\documentclass [12pt]{article}
\def\maj#1{\ifmmode\mbox{\usefont{U}{msb}{m}{n}#1}\else{\usefont{U}{msb}{m}{n}#1}\fi}

\begin{document}

\title{\textbf{Commutation
technique for an exciton photocreated close to
a metal}}
\author{M. Combescot and O. Betbeder-Matibet
 \\ \small{\textit{GPS, Universit\'e Denis Diderot
and Universit\'e Pierre et Marie Curie,
CNRS,}}\\ \small{\textit{Tour 23, 2 place Jussieu, 75251
Paris Cedex 05, France}}}
\date{}
\maketitle

\begin{abstract}
Recently, we have derived the changes in the
absorption spectrum of an exciton when this
exciton is photocreated close to a metal. The
resolution of this problem -- which has similarities
with Fermi edge singularities -- has been made possible
by the introduction of ``exciton diagrams''. The
validity of this procedure relied on a dreadful
calculation based on standard free electron and free
hole diagrams, with the semiconductor-metal interaction
included at second order only, and its intuitive
extention to higher orders. Using the commutation
technique we recently introduced to deal with
interacting excitons, we are now able to \emph{prove}
that this exciton diagram procedure is indeed valid at
any order in the interaction.

\end{abstract}

\vspace{2cm}

			   
\hspace{1.2cm}	71.35.-y -- Excitons and related
phenomena.

\newpage

Interactions with excitons have always been a tricky
problem to handle properly. The interactions being in
fact interactions with free electrons and free holes,
one a priori has to crack the excitons into electrons
and holes, in order to really know their effects. This
leads to see the exciton as the sum of ladder diagrams
$^{(1)}$ between one electron and one hole, with
possibly, once in a while, an interaction of this
electron or this hole with something else. Although
fully safe, this approach becomes very fast dreadfully
complicated, as can be seen from the simplest problem
on interacting excitons studied in reference (2),
namely an exciton photocreated close to a metallic
``mirror''. It is indeed the simplest problem on
interacting excitons, in the sense that the
photocreated electron and the metal electrons are
discernable (being spatially separated) so that there
is no Pauli exclusion between them. This Pauli
exclusion, and the exchange processes associated to the
indiscernability of the carriers, is an additional, but
major, difficulty for interacting exciton problems.
Rather recently, we have developed a ``commutation
technique'' $^{(3,4)}$ which allows to cleanly identify
contributions coming from Coulomb interaction
\emph{between excitons} and contributions coming from
possible exchange \emph{between carriers}. Using this
commutation technique, we can derive the correlations
between excitons at any order \emph{exactly}. We have
already been able to prove that the effective bosonic
hamiltonian for excitons quoted by everyone up to
now cannot be correct : First, it is not even hermitian
\nolinebreak
$^{(3)}$ ; second, it misses purely Pauli terms
$^{(3)}$ ; third, and worse, the concept of effective
hamiltonian itself has to be given up $^{(5)}$ because,
whatever the exciton-exciton part is, it cannot
reproduce the exciton correlations correctly, due to the
complexity of the exchange processes. If such an
effective hamiltonian were correct, exciton diagrams
could obviously be used, with boson-exciton propagators
and interaction vertices deduced from the interacting
part of the hamiltonian. Since such an effective
hamiltonian is incorrect, the validity of the exciton
diagram procedure is actually not established at all.

At the time we studied the problem of an exciton 
photocreated close to a metal and the changes in the 
exciton absorption spectrum induced by the 
semiconductor-metal interaction, we had not yet
developed this commutation technique. This is why we
safely used  standard diagrams $^{(6)}$ with free
electrons and free holes and Coulomb interactions
between them. We  were able to put the electron-metal
and hole-metal  interactions at second order only. At
this order, we proved that the sum of all the seven
complicated  diagrams corresponding to these second
order processes ends up with the same result as the one
derived in an extremely simple way, by using intuitive
``exciton  diagrams'' : In these, the exciton
propagator was taken  to be
\begin{equation}
G_x(\omega;\nu,\mathbf{Q})=\frac{1}{\omega-E_
{\nu,\mathbf{Q}}+i\eta}\ ,
\end{equation}
where
$E_{\nu,\mathbf{Q}}=\varepsilon_\nu
+\mathcal{E}_\mathbf{Q}$ is the energy of the $(\nu,
\mathbf{Q})$ exciton, $\nu$ being the relative motion
state index and $\mathbf{Q}$ the center of mass
momentum. The  exciton-metal vertex was
somehow
\emph{cooked in a  reasonable way} from the bare
electron-metal and  hole-metal interactions.

Since there were no hope to calculate standard
electron-hole diagrams with more than two electron-metal
and hole-metal interactions, we assumed that the exciton 
diagram procedure, which looked physically quite 
reasonable, should hold at any order.

By studying this problem in the light of our commutation 
technique, we
are now able to
\emph{prove}
that this
exciton
diagram procedure is
indeed fully
correct.

Let us
reconsider this
problem from
the beginning :
A
highly doped 2D
quantum well
is set at a
distance $d$ from
an empty
quantum well in
which an
exciton is
photocreated.
The metal Fermi
sea reacts to
the sudden
appearance of
the
photocreated
electron-hole, and its
change, in
turn, modifies
the photon
absorption. Of
course,
similarities 
with Fermi edge
singularities $^{(7-9)}$
follow from
this Fermi sea
reaction.

The hamiltonian
of this
semiconductor-metal
coupled system
reads $H=H_{\mathrm{sc}}+H_\mathrm{m}+W_\mathrm{sc-m}$,
where
$H_\mathrm{sc}$
is the
semiconductor
hamiltonian and
$H_\mathrm{m}$
is the metal
hamiltonian. 
The
semiconductor-metal
coupling $W_\mathrm{sc-m}$ reads
\begin{equation}
W_\mathrm{sc-m}=\sum_{\mathbf{q}\neq\mathbf{0}}
\sum_\mathbf{k}V(\mathbf{q})\left(a_\mathbf{k+q}^\dag
a_\mathbf{k} -b_\mathbf{k+q}^\dag
b_\mathbf{k}\right)w_{-\mathbf{q}}\ ,
\end{equation} 
\begin{equation}
w_\mathbf{q}=\sum_\mathbf{p}c_\mathbf{p+q}
^\dag c_\mathbf{p}\ ,
\end{equation}
$a_\mathbf{k}^\dag$,
$b_\mathbf{k}^\dag$
and
$c_\mathbf{k}^\dag$
being the
semiconductor
electron,
semiconductor
hole and metal
electron
creation
operators
respectively,
while, for metal
and
semiconductor
$d$ apart
$^{(2)}$,\linebreak
$V(\mathbf{q})=e^{-qd}2\pi
e^2/Sq$.

The absorption
of a photon
$(\Omega,\mathbf{Q})$,
given by the
Fermi golden
rule, is
proportional
to the
imaginary part
of the response
function
\begin{equation}
S(\Omega,\mathbf{Q})=\langle i|U\frac{1}{\Omega+
\maj{E}_0-H+i\eta}U^\dag|i\rangle\ ,
\end{equation}   
where $|i\rangle=|v\rangle\otimes|0\rangle$, 
with $|v\rangle$
being the
semiconductor
vacuum state and
$|0\rangle$ the
metal ground state,
$(H_\mathrm{m}-\maj{E}_0)|0\rangle=0$.

The excitons
(i.\ e.\ all bound
and extended
one-pair
eigenstates of the
semiconductor
hamiltonian,
$(H_\mathrm{sc}-E_{\nu,\mathbf{Q}})B_{\nu,\mathbf{Q}}
^\dag|v\rangle=0$)
are related to the
free pairs by
\begin{equation}
B_{\nu,\mathbf{Q}}^\dag=\sum_\mathbf{k}\langle
\mathbf{k}|x_{\nu}
\rangle\,a_{\mathbf{k}+\alpha_e\mathbf{Q}}^\dag\,
b_{-\mathbf{k}+\alpha_h\mathbf{Q}}^\dag\ ,
\end{equation}
\begin{equation}
a_{\mathbf{k}_e}^\dag\,b_{\mathbf{k}_h}^\dag=\sum_\nu
\langle x_\nu|\alpha_h\mathbf{k}_e-\alpha_e\mathbf{k}_h
\rangle\,B_{\nu,\mathbf{k}_e+\mathbf{k}_h}^\dag\ ,
\end{equation} 
where $\alpha_e=1-\alpha_h=m_e/(m_e+m_h)$, $m_e$ and
$m_h$ being the electron and hole masses. Using eq.\
(6), the
semiconductor-photon
interaction reads
\begin{equation}
U^\dag=A\sum_\mathbf{k}a_{\mathbf{k}+\mathbf{Q}}
^\dag\,b_{-\mathbf
{k}}^\dag=A\sum_{\nu,\mathbf{k'}}\langle
x_\nu|
\mathbf{k'}\rangle\,B_{\nu,\mathbf{Q}}
^\dag=A\sum_\nu B_{\nu,\mathbf{Q}}^\dag\langle x_\nu|
\mathbf{r}=\mathbf{0}\rangle\ ,
\end{equation}                                   
(if we set the sample volume equal to 1). 
The response function thus appears as
\begin{equation}
S(\Omega,\mathbf{Q})=A^2\sum_{\nu,\nu'}\langle\mathbf{r}
=\mathbf{0}|x_{\nu'}\rangle\,S_{\nu'\nu}(\Omega,
\mathbf{Q}) \langle x_\nu|\mathbf{r}=\mathbf{0}\rangle\
,
\end{equation}
\begin{equation}
S_{\nu'\nu}(\Omega,\mathbf{Q})=\langle i 
|B_{\nu',\mathbf{Q}}\,\frac{1}{a-H}
\,B_{\nu,\mathbf{Q}}^\dag|i
\rangle\ ,\hspace{1.5cm}a=\Omega+\maj{E}_0+i\eta\ .
\end{equation}

In order to calculate $S_{\nu'\nu}(\Omega,\mathbf{Q})$,
we can note that
\begin{equation}
\left[H,B_{\nu,\mathbf{Q}}^\dag\right]=
\left[H_\mathrm{sc},B_{\nu,\mathbf{Q}}^\dag
\right]+\left[W_\mathrm{sc-m},B_{\nu,\mathbf{Q}}^\dag
\right]=(E_{\nu,\mathbf{Q}}B_{\nu,\mathbf{Q}}
^\dag+V_{\nu,\mathbf{Q}}^\dag)+W_{\nu,\mathbf{Q}}^\dag\
.
\end{equation}
The first commutator, calculated in reference (3),
shows that $V_{\nu,\mathbf{Q}}^\dag$ acts on
semiconductor electron-hole pairs only so that 
$V_{\nu,\mathbf{Q}}^\dag|v\rangle=0$. Using eqs.\
(2,5,6), the second commutator gives
\begin{equation}
W_{\nu,\mathbf{Q}}^\dag=\sum_{\mathbf{q}\neq\mathbf{0},
\nu'}\hat{V}_{\nu'\nu}(\mathbf{q})\,
B_{\nu',\mathbf{Q}+\mathbf{q}}^\dag\,
w_{-\mathbf{q}}\ ,
\end{equation}
\begin{equation}
\hat{V}_{\nu'\nu}(\mathbf{q})=\langle x_{\nu'}|
V(\mathbf{q})(e^{i\alpha_h\mathbf{q}.\mathbf{r}}-
e^{-i\alpha_e\mathbf{q}.\mathbf{r}})|x_\nu\rangle=
\langle x_{\nu'}|\hat{V}(\mathbf{q})|x_\nu\rangle\ .
\end{equation}
$W_{\nu,\mathbf{Q}}^\dag$ physically
corresponds to excite one exciton from a
$(\nu,\mathbf{Q})$ state to a
$(\nu',\mathbf{Q}+\mathbf{q})$ state, whereas the
metal has one of its electrons excited from
$\mathbf{p}$ to $\mathbf{p-q}$.

It is easy to check that eq.\ (10) leads to
\begin{equation}
\frac{1}{a-H}\,B_{\nu,\mathbf{Q}}^\dag=B_{\nu,\mathbf{Q}}
^\dag\,\frac{1}{a-H-E_{\nu,\mathbf{Q}}}+\frac{1}{a-H}\,
(V_{\nu,\mathbf{Q}}^\dag+W_{\nu,\mathbf{Q}}^\dag)\,
\frac{1}{a-H-E_{\nu,\mathbf{Q}}}\
.
\end{equation}
As $V_{\nu,\mathbf{Q}}^\dag|v\rangle=0$, while
$W_{\nu,\mathbf{Q}}^\dag|v\rangle$ writes in terms of
$B_{\nu',\mathbf{Q'}}^\dag$, the iteration of the above
equation allows to generate the expansion of
$S_{\nu'\nu}(\Omega,\mathbf{Q})$ in the exciton-metal
interaction : 
\begin{equation}
S_{\nu'\nu}(\Omega,\mathbf{Q})=\sum_{n=0}^\infty S_{
\nu'\nu}^{(n)}(\Omega,\mathbf{Q})\ .
\end{equation}
The zero order term simply
comes from the first term of eq.\ (13). It reads
\begin{equation}
S_{\nu'\nu}^{(0)}(\Omega,\mathbf{Q})=\langle
i |B_{\nu',\mathbf{Q}}\,B_{\nu,\mathbf{Q}}^\dag\,
\frac{1}{a-H-E_{\nu,\mathbf{Q}}}\,|i\rangle
=\delta_{\nu',\nu}
G_x(\Omega;\nu,\mathbf{Q})\ .
\end{equation}

The first order term appears as
\begin{equation}
S_{\nu'\nu}^{(1)}(\Omega,\mathbf{Q})=G_x(\Omega;\nu,
\mathbf{Q})\sum_{\mathbf{q}_1\neq
0,\nu_1}\langle i|
B_{\nu',\mathbf{Q}}B_{\nu_1,\mathbf{Q}
+\mathbf{q}_1}^\dag\frac{1}{a-H-
E_{\nu_1,\mathbf{Q}+\mathbf{q}_1}}\hat{W}_
{-\mathbf{q}_1;\nu_1\nu}|i\rangle\ ,
\end{equation}
where we have set
$\hat{W}_{-\mathbf{q};\nu'\nu}=\hat{V}_{\nu'\nu}
(\mathbf{q})w_{-\mathbf{q}}$. As
$\langle v|B_{\nu',\mathbf{Q}}B_{\nu_1,\mathbf{Q}+
\mathbf{q}_1}^\dag|v\rangle=\delta_{\nu',\nu_1}
\delta_{\mathbf{q}_1,\mathbf{0}}$, this first order term
is equal to zero.

The second order term reads
\begin{eqnarray}
S_{\nu'\nu}^{(2)}(\Omega,\mathbf{Q})=G_x(\Omega;\nu,
\mathbf{Q})\sum_{\mathbf{q}_2\neq\mathbf{0},\nu_2}
\sum_{\mathbf{q}_1\neq\mathbf{0},\nu_1}
\langle i|B_{\nu',\mathbf{Q}}
B_{\nu_2,\mathbf{Q}+\mathbf{q}_1+\mathbf{q}_2}^\dag
\hspace{4cm}\nonumber\\ \times
\frac{1}{a-H-E_{\nu_2,\mathbf{Q}+
\mathbf{q}_1+\mathbf{q}_2}}\,\hat{W}_{-\mathbf{q}_2;
\nu_2\nu_1}\,
\frac{1}{a-H-E_{\nu_1,
\mathbf{Q}+\mathbf{q}_1}}\,\hat{W}_{-\mathbf{q}_1;
\nu_1\nu}|i\rangle\ .
\end{eqnarray}
The above matrix element can be split into a
semiconductor part and a metal part. The first one
imposes
$\nu_2=\nu'$ and $\mathbf{q}_1+\mathbf{q}_2=0$, so that
\begin{equation}
S_{\nu'\nu}^{(2)}(\Omega,\mathbf{Q})=G_x(\Omega;\nu',
\mathbf{Q})\ T_{\nu'\nu}^{(2)}(\Omega,\mathbf{Q})\,
G_x(\Omega;\nu,\mathbf{Q})\ ,
\end{equation}
\begin{equation}
T_{\nu'\nu}^{(2)}(\Omega,\mathbf{Q})=\sum_{\mathbf{q}_1
\neq 0,\nu_1}\hat{V}_{\nu'\nu_1}(-\mathbf{q}_1)
\langle 0|w_{\mathbf{q}_1}\frac{1}{a
-H_\mathrm{m}-E_{\nu_1,\mathbf{Q}+\mathbf{q}_1}}w_{-\mathbf{q}
_1}| 0\rangle\hat{V}_{\nu_1\nu}(\mathbf{q}_1)\ .
\end{equation}
If we neglect Coulomb interaction between metal
electrons for simplicity, as in reference (2), 
$H_\mathrm{m}\,c_{\mathbf{p}-\mathbf{q}}^\dag
c_\mathbf{p}|0
\rangle=(\maj{E}_0+\epsilon_\mathbf{p-q}-\epsilon
_\mathbf{p})\,c_{\mathbf{p}-\mathbf{q}}^\dag 
c_\mathbf{p}|0\rangle$, $\epsilon_\mathbf{p}$ being the
metal-electron energy. The matrix element of eq.\ (19)
is thus equal to
\begin{equation}
\sum_{|\mathbf{p}|<k_\mathrm{F}<|\mathbf{p}-\mathbf{q}_1|
}\frac{1}{\Omega-(E_{\nu_1,\mathbf{Q}+
\mathbf{q}_1}+\epsilon_{\mathbf{p}-\mathbf{q}_1}-
\epsilon_\mathbf{p})+i\eta}\ .
\end{equation}
We can split it into contributions from the exciton,
the metal electron and the metal hole by
using the standard trick,
\begin{equation}
\frac{1}{\Omega-a-b+i\eta}=\int\frac{id\omega}{2\pi}\,
\left(\frac{1}{\omega+\Omega-a+i\eta}\right)\,\left(
\frac{1}{-\omega-b+i\eta}\right)\ .
\end{equation}
Equation (20) then reads
\begin{equation}
\int\frac{id\omega_1}{2\pi}\,G_x(\omega_1+\Omega;\nu_1,
\mathbf{Q}+\mathbf{q}_1)\left[-\sum_\mathbf{p}\int
\frac{id\omega}{2\pi}\,g(\omega,\mathbf{p})\,g(\omega-
\omega_1,\mathbf{p}-\mathbf{q}_1)\right]\ ,
\end{equation}
where
$g(\omega,\mathbf{p})=\left(\omega-\epsilon_\mathbf{p}
+i\eta\,
\mathrm{sign}(\epsilon_\mathbf{p}-\mu)\right)^{-1}$ is
the usual metal-electron Green's function, while $G_x
(\omega;\nu,\mathbf{Q})$ is the ``exciton Green's
function'' given in eq.\ (1). This leads to rewrite
$T_{\nu'\nu}^{(2)}(\Omega,\mathbf{Q})$ as
\begin{equation}
T_{\nu'\nu}^{(2)}(\Omega,\mathbf{Q})=\sum_{\mathbf{q}_1
\neq 0,\nu_1}\int\frac{id\omega_1}{2\pi}\,B(\omega_1,
\mathbf{q}_1)\,\left[\hat{V}_{\nu'\nu_1}(-\mathbf{q}_1)
\,G_x(\Omega+\omega_1;\nu_1,\mathbf{Q}+\mathbf{q}_1)\,
\hat{V}_{\nu_1\nu}(\mathbf{q}_1)\right]\ ,
\end{equation}
$B(\omega_1,\mathbf{q_1})$ being the standard ``bubble''
contribution as given by the bracket of eq.\ (22). This
response function second order term, as well as the
zero order term given in eq.\ (15), correspond to the
exciton diagrams shown in fig.\ (1), with the
exciton-metal vertex being
$\hat{V}_{\nu'\nu}(\mathbf{q})$.

More generally, the
$n^{\mathrm{th}}$ order term of $S_{\nu'\nu}(\Omega,
\mathbf{Q})$ appears as
\begin{eqnarray}
S_{\nu'\nu}^{(n)}(\Omega,\mathbf{Q})&=&G_x(\Omega;\nu',
\mathbf{Q})\,G_x(\Omega;\nu,\mathbf{Q})\hspace{7cm}
\nonumber\\
&\times&\left[\sum_{\mathbf{q}_{n-1}\neq
\mathbf{0},
\nu_{n-1}}\cdots\sum_{\mathbf{q}_1\neq
\mathbf{0},\nu_1}\langle
0|\hat{W}_{\mathbf{q}_{n-1}+\cdots+\mathbf{q}_1;\nu'
\nu_{n-1}}\,M_{n-1}\,M_{n-2}\cdots M_1|0\rangle
\right]\ ,\nonumber
\\ M_m&=&
\frac{1}{a-H_\mathrm{m}-E_{\nu_m,\mathbf{Q}+
\mathbf{q}_m+\cdots+\mathbf{q}_1}}\,
\hat{W}_{-\mathbf{q}_m;\nu_m\nu_{m-1}}\hspace{2cm}
(\nu_0\equiv\nu)\ .\hspace{1cm}
\end{eqnarray}
The bracket corresponds to all the possible ways to
start with a $(\nu,\mathbf{Q})$ exciton, to excite it
into various $(\nu'',\mathbf{Q}+\mathbf{q}'')$ states
while shaking up the metal Fermi sea by
$(-\mathbf{q}'')$ and to end with a $(\nu',\mathbf{Q})$
exciton. As an example, the $4^\mathrm{th}$ order terms
are shown in fig.\ (2). They are basically of two types
: The first term (fig.\ (2a)) corresponds to excite and
recombine one electron-hole pair in the metal Fermi sea,
twice. Its contribution to $S_{\nu'\nu}^{(4)}(\Omega,
\mathbf{Q})$ is given by
\begin{equation}
G_x(\Omega;\nu',\mathbf{Q})\left[\sum_{\nu_1}
T_{\nu'\nu_1}^{(2)}(\Omega,\mathbf{Q})\,G_x(\Omega;\nu_1,
\mathbf{Q})\,T_{\nu_1\nu}^{(2)}(\Omega,\mathbf{Q})
\right]G_x(\Omega;\nu,\mathbf{Q})\ .
\end{equation}
The other terms of fig.\ (2) can be formally written as
\begin{equation}
G_x(\Omega;\nu',\mathbf{Q})\,T_{\nu'\nu}^{(4)}(\Omega,
\mathbf{Q})\,G_x(\Omega;\nu,\mathbf{Q})\ ,
\end{equation}
where $T_{\nu'\nu}^{(4)}(\Omega,\mathbf{Q})$
corresponds to the transfer of the $(\nu,\mathbf{Q})$
exciton into the $(\nu',\mathbf{Q})$ state associated
to all possible \emph{connected} excitation processes
of the metal Fermi sea with 4 semiconductor-metal
interactions.

This shows that the sum of all contributions to 
$S_{\nu'\nu}(\Omega,\mathbf{Q})$ reads
\begin{eqnarray}
S_{\nu'\nu}(\Omega,\mathbf{Q})=\delta_{\nu',\nu}\,G_x
(\Omega;\nu,\mathbf{Q})\hspace{10cm}\nonumber
\\ +G_x(\Omega;\nu',\mathbf{Q})\left[T_{\nu'\nu}(
\Omega,\mathbf{Q})+\sum_{\nu_1}T_{\nu'\nu_1}(\Omega,
\mathbf{Q})G_x(\Omega;\nu_1,\mathbf{Q})T_{\nu_1\nu}
(\Omega,\mathbf{Q})+\cdots\right]G_x(\Omega;\nu,\mathbf
{Q})\ ,
\end{eqnarray}
where $T_{\nu'\nu}(\Omega,\mathbf{Q})$ corresponds to
the transfer of a $(\nu,\mathbf{Q})$ exciton into a 
$(\nu',\mathbf{Q})$ state associated to the sum of all
possible \emph{connected} excitation processes of the
metal Fermi sea with two or more semiconductor-metal
interactions. This expansion of $S_{\nu'\nu}(\Omega,
\mathbf{Q})$ is shown on fig.\ (3). It corresponds to
the expansion of the integral equation shown in fig.\
(3).

It is in fact possible to rewrite $S_{\nu'\nu}(\Omega,
\mathbf{Q})$, as well as $S(\Omega,\mathbf{Q})$, in a
quite compact form : For that, we first rewrite the
exciton propagator as
\begin{equation}
G_x(\Omega;\nu,\mathbf{Q})=\langle x_\nu|\frac{1}
{\Omega-h_x-\mathcal{E}_\mathbf{Q}+i\eta}|x_\nu\rangle\
,
\end{equation}
where $h_x$ is the exciton relative motion hamiltonian,
$(h_x-\varepsilon_\nu)|x_\nu\rangle=0$. By noting that
the second order transfer, given in eq.\ (23), also
reads
$T_{\nu'\nu}^{(2)}(\Omega,\mathbf{Q})=\langle x_{\nu'}|
T^{(2)}(\Omega,\mathbf{Q})|x_\nu\rangle$ with
\begin{equation}
T^{(2)}(\Omega,\mathbf{Q})=\sum_{\mathbf{q}\neq\mathbf
{0}}\int\frac{id\omega}{2\pi}\,B(\omega,\mathbf{q})\,
\hat{V}({-\mathbf{q}})\,\frac{1}{\Omega+\omega-h_x-
\mathcal{E}_\mathbf{Q+q}+i\eta}\,\hat{V}(\mathbf{q})\ ,
\end{equation}
we can, in a similar way, rewrite the higher order
transfers as $T_{\nu'\nu}(\Omega,\mathbf{Q})=
\langle x_{\nu'}|T(\Omega,\mathbf{Q})|x_\nu\rangle$.
Since $\delta_{\nu',\nu}=\langle
x_{\nu'}|x_\nu\rangle$, eq.\ (27) is nothing but
the expansion of
\begin{equation}
S_{\nu'\nu}(\Omega,\mathbf{Q})=\langle x_{\nu'}|
\frac{1}{\Omega-h_x-T(\Omega,\mathbf{Q})-\mathcal{E}
_\mathbf{Q}
+i\eta}|x_\nu\rangle\ ,
\end{equation}
so that the response function $S(\Omega,\mathbf{Q})$
given in eq.\ (8) takes the quite compact form,
\begin{equation}
S(\Omega,\mathbf{Q})=A^2\langle\mathbf{r}=\mathbf{0}|
\frac{1}{\Omega-h_x-T(\Omega,\mathbf{Q})-\mathcal{E}
_\mathbf{Q}
+i\eta}|\mathbf{r}=\mathbf{0}\rangle\ .
\end{equation}
The above equation is exactly the eq.\ (9) of reference
(1). The explicit form of this response function was
then obtained in terms of the right and left eigenstates
$|\hat{x}_\nu\rangle$ and $|\hat{\hat{x}}_\nu\rangle$
of the non hermitian ``hamiltonian''
$h_x+T(\Omega,\mathbf{Q})$. As its eigenvalues are
complex, the exciton absorption lines in the presence
of a 2D metal have now tails.

\emph{In conclusion}, our commutation technique allows
to prove in a quite transparent way that the problem of
the exciton absorption spectrum changes induced by
the presence of a distant metal, can indeed be solved
within exciton diagrams at any order in the
semiconductor-metal interaction. These diagrams
visualize the fact that a $(\nu,\mathbf{Q})$ exciton
is created by a $\mathbf{Q}$ photon. This 
$(\nu,\mathbf{Q})$ exciton scatters to a
$(\nu_1,\mathbf{Q}+\mathbf{q}_1)$ state and then to a 
$(\nu_2,\mathbf{Q}+\mathbf{q}_1+\mathbf{q}_2)$ state and
so on \ldots At each $\mathbf{q}_i$ scattering, a
$(-\mathbf{q}_i)$ metal electron-metal hole pair is
excited. The photocreated exciton must end all these
scatterings in a
$(\nu',\mathbf{Q})$ state in order to possibly recombine
into a $\mathbf{Q}$ photon. On a technical point of
view, to each $(\nu',\mathbf{Q}')$ exciton we
associate the propagator 
$G_x(\omega;\nu',\mathbf{Q}')$ given in eq.\ (1).
To each scattering of a
$(\nu,\mathbf{Q};\mathbf{p})$ exciton-metal-electron
state into a $(\nu',\mathbf{Q+q};\mathbf{p-q})$ state
we associate the exciton-metal vertex
$\hat{V}_{\nu'\nu}(\mathbf{q})$, given in eq.\ (12),
and we conserve
$\omega$ and
$\mathbf{q}$ at each vertex, as usual
for diagrams.

\newpage

\hbox to \hsize {\hfill REFERENCES
\hfill}

\vspace{0.5cm}

\noindent
(1) A. Fetter, J. Walecka, \emph{Quantum Theory of
Many-Particle Systems} (McGraw-Hill, New York, 1971).

\noindent
(2) M. Combescot, O. Betbeder-Matibet, B. Roulet,
Europhys.\ Lett.\ \underline{57}, 717 (2002).

\noindent
(3) M. Combescot, O. Betbeder-Matibet, Europhys.\
Lett.\ \underline{58}, 87 (2002).

\noindent
(4) O. Betbeder-Matibet, M. Combescot, Eur.\ Phys.\ J.\
B \underline{27}, 505 (2002).

\noindent
(5) M. Combescot, O. Betbeder-Matibet, Cond-mat/0201554
; to appear in Europhys.\ Lett.\ .

\noindent
(6) O. Betbeder-Matibet, M. Combescot, Eur.\ Phys.\ J.\
B \underline{22}, 17 (2001).

\noindent
(7) G. Mahan, Phys.\ Rev.\ \underline{163}, 612 (1967).

\noindent
(8) P. Nozieres, C. T. De Dominicis, Phys.\ Rev.\
\underline{178}, 1097 (1969).

\noindent
(9) M. Combescot, P. Nozieres, J.\ Phys.\ (Paris)
\underline{32}, 913 (1971).

\newpage

\hbox to \hsize {\hfill FIGURE CAPTIONS
\hfill}
\vspace{0.5cm}

Fig.\ (1) :

Response function in terms of exciton diagrams, at zero
order (a) and at second order (b) in the
semiconductor-metal interaction.

To the  $(\Omega;\nu,\mathbf{Q})$ exciton, we associate
the exciton propagator $G_x(\Omega;\nu,\mathbf{Q})$
given in eq.\ (1), and to the scattering of a
$(\Omega;\nu,\mathbf{Q})$ exciton-$(\omega,\mathbf{p})$
metal electron into a $(\Omega+\omega_1;\nu_1,\mathbf{Q}
+\mathbf{q}_1)$ exciton-$(\omega-\omega_1,\mathbf{p}-
\mathbf{q}_1)$ metal electron, we associate the
exciton-metal vertex $\hat{V}_{\nu_1\nu}(\mathbf{q}_1)$
given in eq.\ (12).

Fig.\ (2) :

Exciton diagrams for the response function at
$4^{\mathrm{th}}$ order in the semiconductor-metal
interaction.

Fig.\ (3) :

Integral equation verified by
$S_{\nu'\nu}(\Omega,\mathbf{Q})$ as given in eq.\ (27).

\end{document}